\documentclass[conference]{IEEEtran}
\IEEEoverridecommandlockouts
\usepackage{cite}
\usepackage{amsmath,amssymb,amsfonts}
\usepackage{algorithm}
\usepackage{algorithmic}
\usepackage{booktabs}
\usepackage{graphicx}
\usepackage{subcaption}
\usepackage{textcomp}
\usepackage{xcolor}
\usepackage{multirow}
\usepackage{url}
\usepackage{flushend}
\def\BibTeX{{\rm B\kern-.05em{\sc i\kern-.025em b}\kern-.08em
    T\kern-.1667em\lower.7ex\hbox{E}\kern-.125emX}}
\begin{document}

\title{Yet Another Watermark for Large Language Models}

\author{Siyuan Bao$^1$, Ying Shi$^1$, Zhiguang Yang$^1$, Hanzhou Wu$^{1,2,\star}$\thanks{$^\star$\emph{Corresponding author: Hanzhou Wu (contact email: h.wu.phd@ieee.org)}} and Xinpeng Zhang$^1$\\
$^1$School of Communication and Information Engineering, Shanghai University, Shanghai 200444, China\\
$^2$School of Big Data and Computer Science, Guizhou Normal University, Guiyang 550025, China
}

\maketitle

\begin{abstract}
Existing watermarking methods for large language models (LLMs) mainly embed watermark by adjusting the token sampling prediction or post-processing, lacking intrinsic coupling with LLMs, which may significantly reduce the semantic quality of the generated marked texts. Traditional watermarking methods based on training or fine-tuning may be extendable to LLMs. However, most of them are limited to the white-box scenario, or very time-consuming due to the massive parameters of LLMs. In this paper, we present a new watermarking framework for LLMs, where the watermark is embedded into the LLM by manipulating the internal parameters of the LLM, and can be extracted from the generated text without accessing the LLM. Comparing with related methods, the proposed method entangles the watermark with the intrinsic parameters of the LLM, which better balances the robustness and imperceptibility of the watermark. Moreover, the proposed method enables us to extract the watermark under the black-box scenario, which is computationally efficient for use. Experimental results have also verified the feasibility, superiority and practicality. This work provides a new perspective different from mainstream works, which may shed light on future research.
\end{abstract}

\begin{IEEEkeywords}
Watermarking, large language models, intellectual property protection, parameter modulation.
\end{IEEEkeywords}

\section{Introduction}
Generative artificial intelligence, represented by, e.g., ChatGPT and DeepSeek, is widely used for generative tasks such as text composition, image creation, and speech synthesis, which greatly enhances the efficiency and quality of digital content production. However, while generative content brings innovative opportunities, it also gives rise to a series of security and ethical issues, including the spread of misinformation, misuse of deepfakes, difficulty in identifying content authenticity, and unclear copyright ownership, all of which pose serious threats to social network security. As generative artificial intelligence advances rapidly, effectively identifying its generated content is of significant practical importance, which makes watermarking for generative artificial intelligence, also called \emph{generative model watermarking}, a hot research topic in recent years.

Early watermarking methods are designed for conventional generative models \cite{Wu:TCSVT:2021, Zhang:IH:2024, Liu:IoT:2024, Zhang:IFDMC:2022}, namely, the number of parameters in the model is relatively small. These watermarking methods are typically realized by retraining or fine-tuning the generative model, which works for small models but cannot be extended to large models that contain a massive number of parameters since retraining or fine-tuning a large model is time consuming and requires a large amount of computing power. Although the idea of marking the generated content is very desirable for both small and large models, how to realize it in large models in an efficient way is very challenging, especially for large language models (LLMs), which generate texts as output and have been widely applied in various application scenarios. 

Currently, there are two common strategies for LLM watermarking. One treats LLM and watermark as two independent functional modules. That is, the watermark can be embedded into the text generated by an LLM in a plug-and-play or post-processing manner. Its advantages include flexibility and high computational efficiency. Since the watermark is independent of the text generation mechanism itself, this strategy, however, are susceptible to interference from noise or external manipulations, reducing the detection accuracy of the watermark. To insert the watermark, it may be necessary to replace words or adjust sentence structures, which can make the text less natural or slightly alter its meaning, impairing imperceptibility.

The other strategy alters the sampling probabilities of candidate tokens, increasing the frequency of tokens from a specific set in the generated text to achieve watermark embedding \cite{Kirchenbauer:ICML:2023}. This method does not pay attention to the characteristics of model weights, but manipulates the logits in a hard way. This blunt and straightforward strategy can enhance the robustness, however, lacks connection to the text generation dynamics of the model. As a result, the text quality may be degraded. 

In this paper, we introduce a new watermarking framework for LLMs. The idea is to apply structured, sparse, and small-scale manipulations to the selected weights in the output layer, thereby amplifying the logits of designated tokens to embed controllable watermark signals. It enhances the probability of sampling specific words at the level of internal parameters of the LLM. For watermark detection, the verifier only requires the knowledge of token selection rule, enabling efficient and reliable verification without access to the model. In summary, the main contributions of the proposed framework include:

\begin{itemize}
\item \emph{Stealthy yet detectable:} Watermarks are embedded at the parameter level rather than the text level, providing strong concealment while ensuring reliable detection.
\item \emph{Text-quality preserving:} Manipulations are subtle and proportionally constrained, thereby ensuring that semantics, style, and fluency remain intact.
\item \emph{Robust and scalable:} The proposed framework generalizes across LLMs of varying sizes and remains effective under common text modifications, ensuring adaptability and stability.
\end{itemize}

The rest structure of this paper is organized as follows. We firstly provide preliminary concepts in Section II, followed by the proposed method in Section III. Experimental results and analysis are provided in Section IV. Finally, we conclude this paper and give discussion in Section V.

\section{Preliminaries}
\subsection{Large Language Models (LLMs)}
LLMs have evolved from traditional statistical, neural, and pretrained language models through the large-scale expansion of parameters, training corpora, and computational resources, enabling them to address increasingly complex tasks. Prominent closed-source systems include the GPT and PaLM series, while open-access models such as LLaMA and DeepSeek have likewise gained significant traction \cite{Zhao:arXiv:2025v16}.

LLMs are built upon the Transformer architecture \cite{Vaswani:NIPS:2017}, which leverages Multi-head Self-attention to effectively capture long-range dependencies in text. Given a token sequence, all tokens are mapped into high-dimensional embeddings, combined with positional encodings, and then processed by stacked encoder layers. Within each layer, self-attention mechanism computes token-level dependencies and assigns attention weights, while feed forward networks (FFNs) with nonlinear activations enhance representation capacity. The hidden states from the final decoder layer are linearly projected into logits. Applying the softmax function to the logits yields a conditional probability distribution over vocabulary. Finally, a decoding strategy such as nucleus sampling \cite{Holtzman:ICLR:2020}, greedy search, temperature sampling, or beam search, is used for next token generation, in an auto-regression fashion.

Pretraining LLMs with self-supervised objectives, e.g., language modeling, denoising autoencoding, or hybrid denoisers, enables LLMs to acquire broad world knowledge. Subsequent instruction tuning aligns the pretrained large model with human instructions, granting it instruction-following capability. With the enormous number of parameters involved, Parameter-Efficient Fine-Tuning (PEFT) methods have been developed to mitigate computational overhead, e.g., \cite{Hu:ICLR:2022}. To mitigate biases and enhance alignment with human preferences, Reinforcement Learning from Human Feedback (RLHF) \cite{Ouyang:NIPS:2022} has been employed. In addition, prompt learning (PL) offers an effective paradigm for adapting pretrained models to downstream tasks.

\subsection{LLM Watermarking}
LLM watermarking has been an essential technique for content attribution, traceability, and copyright protection. Unlike traditional post-processing approaches, watermarking mechanisms that are integrated into the text generation process better preserve the naturalness and diversity of text while embedding reliable noise-like signals. A straightforward idea is to train or fine-tune a model for embedding a watermark into the output, e.g., \cite{Xu:arXiv:2025, Gu:ICLR:2024}, which works but requires sufficient computational power, limiting their applications. Another idea is to refine the sampling strategy for text generation, either at the token level \cite{Dathathri:Nature:2024}  or sentence level \cite{Hou:ACL:2024}, which does not require adjustments to model training, but would increase the inference complexity. An alternative strategy involves manipulating the logits prior to softmax by adding a bias term to the probabilities of specific tokens \cite{Kirchenbauer:ICML:2023}, enhancing the robustness. However, it lacks deep connection to the text generation dynamics of the large model, resulting in that, the text quality may be degraded. It motivates us to introduce a new framework to exploit the text generation dynamics of LLMs, thus achieving better performance. 

\section{Proposed Method}
\subsection{General Framework}
Our core idea is to embed watermark information into LLMs by manipulating model parameters, ensuring that the generated text carries imperceptible watermark signal and allows reliable detection of the watermark signal with a secret key. We argue that focusing on the Transformer’s output linear layer enables tight coupling between watermark and model without altering the main generative pipeline. Based on this, in the embedding phase, the secret key determines a set of tokens whose output weights are slightly amplified, thus biasing generation towards these tokens. For watermark detection, the frequency of these tokens is statistically analyzed to reveal deviations from non-marked distribution, enabling efficient verification.

\subsection{Watermark Embedding}
In LLMs, the Transformer decoder is typically followed by a linear layer processed with softmax, which produces the final prediction probabilities over tokens. The hidden state $\mathbf{h}$ from the last decoder layer encodes rich sequence-level linguistic and semantic features, but remains an abstract and continuous representation. Once passed through the linear layer, $\mathbf{h}$ will be projected into the vocabulary space, thereby generating the logit distribution used for token sampling, i.e.,
\begin{equation}
\mathbf{l} = \mathbf{W}\mathbf{h} + \mathbf{b},
\end{equation}
where $\mathbf{W} \in \mathbb{R}^{|\mathcal{V}| \times d}$ represents the output weight matrix, $|\mathcal{V}|$ is the vocabulary size, and $d$ is the hidden dimension.  

In the autoregressive generation process, the LLM generates tokens sequentially, with the prediction at step $i$ conditioned on the preceding $i-1$ tokens by maximizing $\mathrm{Pr}(t_i|t_1, \dots, t_{i-1})$. The logits $\mathbf{l} = (l_1, l_2, ..., l_{|\mathcal{V}|})^\mathrm{T}$ at step $i$ are transformed into probabilities through the softmax function, i.e.,
\begin{equation}
p_j = \mathrm{softmax}(l_j) = \frac{e^{l_j}}{\sum_{j=1}^{|\mathcal{V}|}e^{l_j}}, \forall 1\leq j\leq |\mathcal{V}|.
\end{equation}
 
The matrix $\mathbf{W}$ and the bias term $\mathbf{b}$ in the output linear layer actually serve as the key parameters in line with the encoder-decoder architecture. They map semantic representations into the vocabulary distribution space, and their parameter scale is comparatively modest. Consequently, applying subtle perturbations to these parameters provides an effective means of embedding watermark signals. Such fine-grained interventions introduce negligible impact on the model’s language modeling capability, thereby preserving the fluency and overall quality of the generated text containing hidden signal.

Let $f(\mathrm{key})$ denote a pseudo-random function controlled by a secret key. We use it for step $i$ to generate a subset of $\mathcal{V}$:
\begin{equation}
\mathcal{G}_i = \mathrm{TokenSelection}\{\mathcal{V},f(\mathrm{key}_i),\gamma\},
\end{equation}
where $|\mathcal{G}_i| = \gamma |\mathcal{V}|$ and $\gamma \in (0,1)$ is the parameter controlling the fraction of tokens selected. For each $g \in \mathcal{G}_i$, its index value is defined as $\mathrm{idx}(g)$, i.e., $p_{\mathrm{idx}(g)}$ corresponds to the prediction probability of $g$. Let $\mathbf{W}[\mathrm{idx}(g),:]$ and $b_{\mathrm{idx}(g)}$ be the $\mathrm{idx}(g)$-th row vector of $\mathbf{W}$ and $\mathrm{idx}(g)$-th component of $\mathbf{b}$, respectively. For those LLMs without the bias term $\mathbf{b}$, we have $\mathbf{b} \equiv \mathbf{0}$. We are to modify $\mathbf{W}$ and/or $\mathbf{b}$ to embed watermark for step $i$.

\begin{algorithm}[!t]
 \caption{Pseudocode for watermark embedding at step $i$}
 \begin{algorithmic}[1]
	\renewcommand{\algorithmicrequire}{\textbf{Input:}}
	\renewcommand{\algorithmicensure}{\textbf{Output:}}
	\REQUIRE Token-sequence $(t_1, t_2, ..., t_{i-1})$, model $\mathcal{M}$, $\mathcal{V}$, $\mathbf{W}$, $\mathbf{b}$, $\mathrm{key}_i$, $\gamma$, $\alpha_\uparrow$, $\alpha_\downarrow$, $\beta_\uparrow$ and $\beta_\downarrow$.
	\ENSURE  Token $t_i$
	\STATE Apply Eq. (3) to determine $\mathcal{G}_i$
	\FOR {each $v \in \mathcal{V}$ }
        \IF { $g \in \mathcal{G}_i$ }
            \STATE Apply Eq. (4)
        \ELSE
            \STATE Apply Eq. (5)
        \ENDIF
	\ENDFOR
    \STATE Determine the token $t_i$ using modified $\mathbf{W}$ and $\mathbf{b}$  
	\RETURN $t_i$
 \end{algorithmic}
\end{algorithm}

Recalling that the method in \cite{Kirchenbauer:ICML:2023} added a hardness parameter to every ``green-list'' logit, its essence is to add the parameter to some components of $\mathbf{b}$, so that ``green-list'' tokens are more likely to be sampled as output. This rigid strategy may ensure robustness but overlooks the model’s text-generation dynamics, potentially degrading output quality and thus diminishing the watermark’s concealment. We propose a new strategy that better balances robustness and imperceptibility. 

Formally, at step $i$, for each $g \in \mathcal{G}_i$, we modify $\mathbf{W}[\mathrm{idx}(g),:]$ and $b_{\mathrm{idx}(g)}$ as:
\begin{equation}
\mathbf{W}[\mathrm{idx}(g),:] = \alpha_\uparrow\mathbf{W}[\mathrm{idx}(g),:]~\mathrm{and}~b_{\mathrm{idx}(g)} = \beta_\uparrow b_{\mathrm{idx}(g)}.
\end{equation}
For each $v \in \mathcal{V}\setminus\mathcal{G}_i$, we modify $\mathbf{W}[\mathrm{idx}(v),:]$ and $b_{\mathrm{idx}(v)}$ as:
\begin{equation}
\mathbf{W}[\mathrm{idx}(v),:] = \alpha_\downarrow\mathbf{W}[\mathrm{idx}(v),:]~\mathrm{and}~b_{\mathrm{idx}(v)} = \beta_\downarrow b_{\mathrm{idx}(v)}.
\end{equation}
Here, $\alpha_\uparrow$, $\beta_\uparrow$, $\alpha_\downarrow$ and $\beta_\downarrow$ are scaling factors satisfying $\alpha_\uparrow \geq 1$, $\beta_\uparrow \geq 1$, $\alpha_\downarrow \leq 1$ and $\beta_\downarrow \leq 1$. Thus, Ref. \cite{Kirchenbauer:ICML:2023} can be roughly considered as a special case of the proposed method, i.e., $\alpha_\uparrow  \equiv 1$, $\beta_\uparrow > 1~(b_{\mathrm{idx}(g)} > 0)$, $\alpha_\downarrow \equiv 1$ and $\beta_\downarrow \equiv 1$. $\mathbf{W}$ projects the hidden vector $\mathbf{h}$ containing rich sequence-level linguistic and semantic features into the token space, providing a profound reflection of the dynamics of large models. Thus, $\alpha_\uparrow$ and $\alpha_\downarrow$ makes the watermark more tightly entangled with the model, while $\beta_\uparrow$ and $\beta_\downarrow$ are approximately independent of the model. 

Intuitively, reasonably setting the values of these parameters can achieve better watermarking performance. Moreover, this embedding strategy requires no retraining or gradient updates. It only adjusts the output layer parameters, ensuring simplicity and efficiency. As this method does not depend on the internal structure of the Transformer backbone, it is broadly applicable across different types and scales of LLMs, thus offering strong generalizability and portability. Furthermore, the localized parameter modification ensures controllable impact, facilitating flexible integration in real-world scenarios. Algorithm 1 shows the pseudocode for watermark embedding at step $i$. 

\subsection{Watermark Extraction}
Although using $\alpha_\uparrow$ and $\alpha_\downarrow$ does not guarantee an increase in the sampling probability of every selected token, we believe that in most cases it is increased, leading to a systematic rise in their frequency across generated text, which forms a detectable signal. During watermark detection, we count the occurrences of selected tokens corresponding to $\mathcal{G}_i$ in the candidate text and compare them with the theoretical distribution under the null hypothesis of ``no watermark''. If the observed deviation reaches statistical significance, the given text is considered to be \emph{marked}. This process can be identical to the method in \cite{Kirchenbauer:ICML:2023}.

\begin{figure}[!t]
    \centering
    \includegraphics[width=\linewidth]{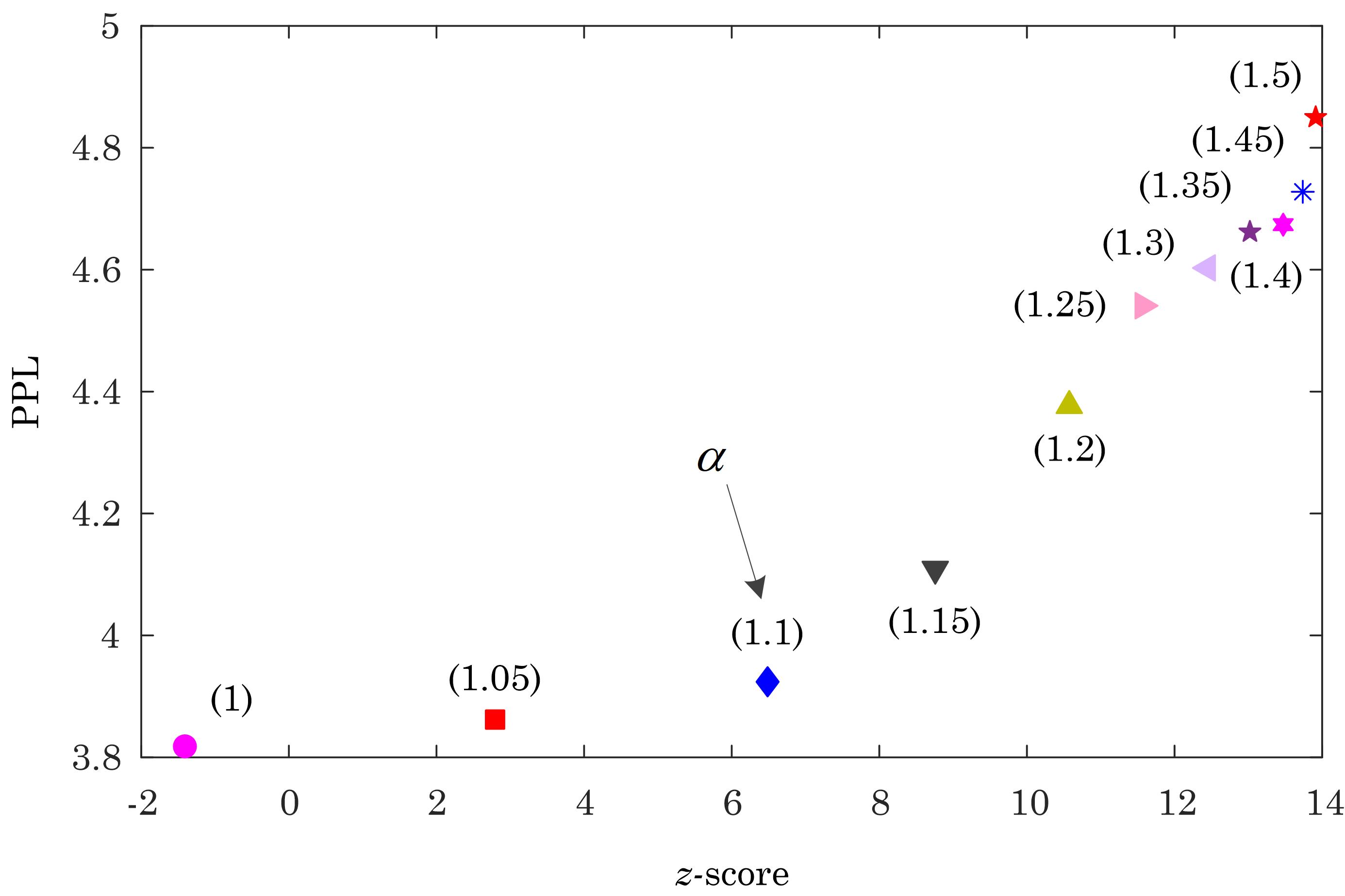}
    \caption{The text quality (PPL) and detection performance ($z$-score) due to different $\alpha$. Here, we use $\gamma = 0.5$.}
    \label{F1}
\end{figure}

\section{Experimental Results and Analysis}
\subsection{Setup}
Although there are different ways to set the scaling factors, we focus on the boundary case to ensure fair comparison with related works. In our experiments, we set $\alpha_\uparrow = \alpha$ and $\alpha_\downarrow = \beta_\uparrow = \beta_\downarrow = 1$. Moreover, we set $\mathcal{G}_i \equiv \mathcal{G}_j$ for all $i\neq j$, i.e., all token positions use the same secret key applied to the function $f$. Accordingly, we can use $z$-score introduced in \cite{Kirchenbauer:ICML:2023} to detect the presence of the watermark in a text.

One of the most popular LLMs, i.e., LLaMA3-8B\footnote{Online available: \url{https://huggingface.co/meta-llama}}, is used for our experiments. The C4 dataset \cite{Dodge:EMNLP:2021} is used for evaluation, from which $400$ natural texts are randomly selected as prompt and fed into the model to generate either watermarked texts or non-watermarked texts. The quality of these generated texts is measured by perplexity (PPL) \cite{Yang:Math:2025}, whereas detection of the watermark relies on the $z$-score statistic. To assess robustness under realistic conditions, we here use three representative text perturbation operations, i.e., masking, deletion, and insertion, and evaluate our method across different attack degrees. It is noted that the larger the $z$-score, the higher the likelihood of the watermark’s presence, e.g., when the $z$-socre is $2.33$, the confidence in the presence of the watermark approaches $99$\%.

\begin{figure}[!t]
    \centering
    \includegraphics[width=\linewidth]{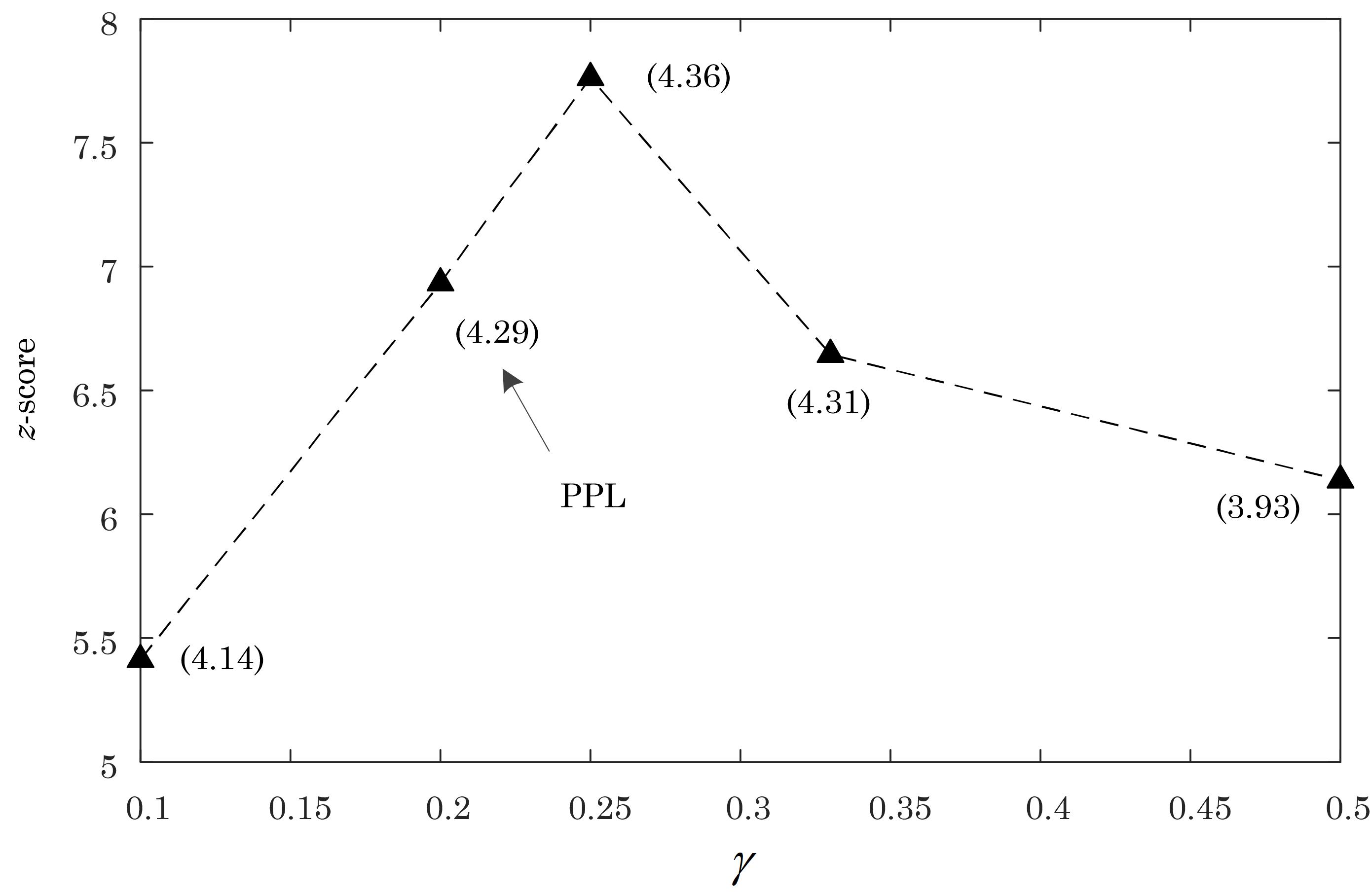}
    \caption{The text quality (PPL) and detection performance ($z$-score) due to different $\gamma$. Here, we use $\alpha = 1.1$.}
    \label{F2}
\end{figure}

\subsection{Results}
A larger $\alpha$ corresponds to a stronger degree of embedding, which means that the watermark is more likely to be detected successfully. As shown in Fig. \ref{F1}, the $z$-score already exceeds $2.33$ at $\alpha = 1.05$, and surpasses $6.48$ at $\alpha = 1.1$, demonstrating a very high level of confidence. It is noted that, $\alpha = 1$ corresponds to the original model without watermarking. For the text quality, the PPL remains generally close to that of the original model, showing only a slight increase as $\alpha$ gradually rises. As $\alpha$ approaches $1.3$, the law of diminishing marginal returns becomes evident: the growth rate of the $z$-score slows markedly, signaling saturation in watermark embedding performance. Beyond this threshold, further resource investment yields diminishing cost-effectiveness.

The parameter $\gamma$ determines the proportion of tokens in the vocabulary to be used for watermarking. As shown in Fig. \ref{F2}, $z$-score first increases with $\gamma$, reaches a near-optimal point, and then declines. A small $\gamma$ indicates that the hidden watermark signal is relatively weak. As $\gamma$ increases, the watermark signal becomes stronger and more reliably detected. However, large $\gamma$ dilutes the statistical distinctiveness of the signal, reducing detectability. As a result, an appropriate value of $\gamma$ should be selected to strike a balance among different evaluation metrics in practice. Regardless, the quality of the text is consistently well maintained, referring to the PPLs shown in Fig. \ref{F2}. 

\begin{figure}[!t]
    \centering
    \includegraphics[width=\linewidth]{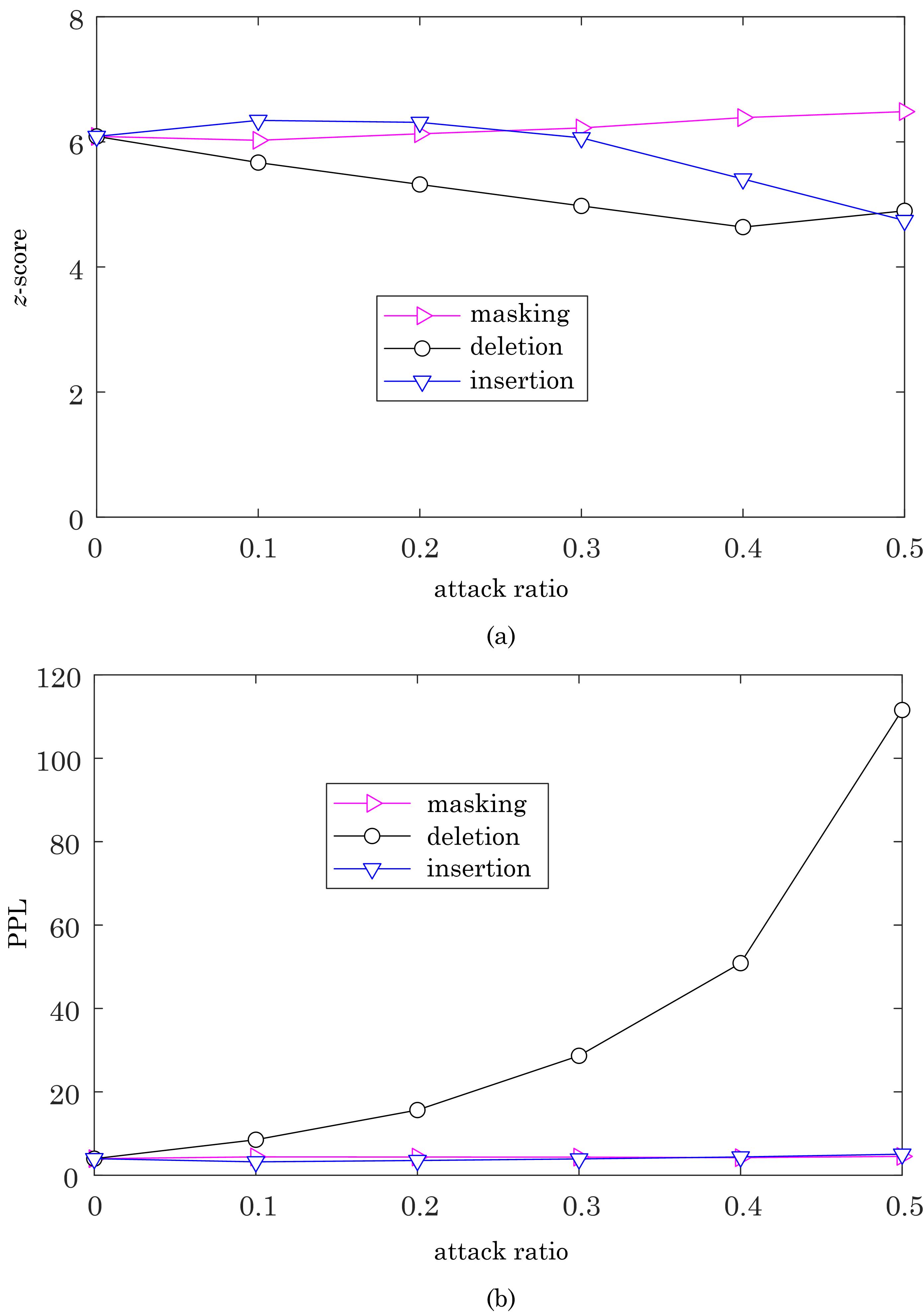}
    \caption{The robustness of the watermark against common text-level attacks. Here, we use $\alpha = 1.1$ and $\gamma = 0.5$.}
    \label{F3}
\end{figure}

We further evaluate the robustness of the hidden watermark against common text-level attacks, including random masking, deletion, and insertion. Fig. \ref{F3} reports $z$-score and PPL due to different degrees of attack. In Fig. \ref{F3}, the attack ratio refers to the proportion of tokens that are randomly masked, deleted, or inserted. It can be seen that the proposed method maintains high $z$-scores even under moderate attacks, confirming that the watermark remains statistically detectable. On the other hand, as the ratio increases, while the PPL due to deletion gradually and reasonably increase, the PPL due to masking and insertion are maintained very well, implying that the proposed method has a significant advantage in maintaining text quality. Overall, our method achieves a very good balance between robustness and imperceptibility of the watermark.

In addition, we compare the proposed method with related works. As shown in Table \ref{tab1}, the proposed method consistently achieves strong detectability of the watermark (i.e., with a high $z$-score) while maintaining a small perplexity compared with the baseline, which has confirmed the superiority. While some related techniques yield a slightly higher $z$-score, they achieve this at the cost of significantly degrading generation quality. 

\begin{table}[!t]
\centering
\caption{Performance comparison between different watermarking methods. Here, we use $\alpha = 1.1$ and $\gamma = 0.5$. Related methods use the default parameter setting in their papers.}
\begin{tabular}{c|cc}
\hline\hline
Method & $z$-score & PPL\\
\hline
Baseline (i.e., non-watermarked) & -1.42 & 3.82 \\
Kirchenbauer \emph{et al.} \cite{Kirchenbauer:ICML:2023}  & 7.34  & 5.08 \\
Lee \emph{et al.} \cite{Lee:ACL:2024}  & 8.20  & 5.05 \\
Lu \emph{et al.} \cite{Lu:ACL:2024}  & 8.50  & 5.03 \\
Dathathri \emph{et al.} \cite{Dathathri:Nature:2024}  & 2.55  & 5.07 \\
Proposed Method  & 6.49  & 3.93 \\
\hline\hline
\end{tabular}\label{tab1}
\end{table}

\section{Conclusion and Discussion}
We introduce a watermarking framework for LLMs, which embeds imperceptible signals into the text generation process to enable content traceability and copyright verification. This method injects sparse, structured modulations into the output-layer weights, ensuring that watermark signals persist throughout generation while remaining invisible in the generated text. Unlike training-based approaches, the proposed framework can be applied to any generative model without retraining. During detection, access to the original model is not required; instead, statistical tests on the distribution of pre-selected tokens suffice to identify watermarked content.  

In experiments, we tuned one scaling factor, i.e., $\alpha_\uparrow$, for fair evaluation. This strategy makes the algorithm computationally efficient. Other advantages include:
\begin{itemize}
    \item \emph{Theoretical interpretability / reparameterization:} The signal is embedded in model parameters and can be regarded as a reparameterization of the final layer, which facilitates theoretical analysis and understanding.
    \item \emph{Distribution smoothness:} The strategy of weight scaling continuously regulates the logits, thereby avoiding abrupt bias at single points and making the generated probability distribution smoother.
    \item \emph{Text quality preservation:} Smooth adjustment of generation probabilities reduces interference with text fluency, resulting in more natural text generation compared with the so-called ``red/green-list'' approach.
\end{itemize}

Experiments show that the proposed method preserves semantic fidelity and fluency across different modulation ratios and allows reliable detection under text-editing operations such as insertion, deletion, and masking, demonstrating robustness.

We want to emphasize that, from a technical point of view, watermarking for LLMs is not new since many previous works have already studied using texts generated by language models to carry additional information, which is essentially embedding watermarks into language models, e.g., \cite{Kang:EI:2020, Guo:JIVP:2021, Yi:SPL:2021, Zheng:SCN:2022, Yang:TDSC:2024}. Our future work focuses on extending these advanced strategies to LLMs.

\section*{Acknowledgment}
This article was partly supported by the Science and Technology Commission of Shanghai Municipality (STCSM) under the Grant Number 24ZR1424000, the National Natural Science Foundation of China (NSFC) under Grant Number U23B2023, 2024 Xizang Autonomous Region Central Guided Local Science and Technology Development Fund Project under Grant Number XZ202401YD0015, and the Basic Research Program for Natural Science of Guizhou Province under Grant Number QIANKEHEJICHU-ZD[2025]-043.


\end{document}